**Quantum teleportation protocol with assistant who prepares the amplitude modulated unknown qubits**

Sergey A. Podoshvedov

Department of computer modeling and nanotechnology, Institute of natural and exact sciences, South Ural State University, Lenin Av. 76, Chelyabinsk, Russia
e-mail: sapodo68@gmail.com

**Abstract** We consider novel implementation of quantum teleportation protocol of unknown qubit with entangled hybrid state. Coherent components of the entangled hybrid state displace the teleported qubit at the same absolute but opposite in sign values, so that any information about amount by which the qubit is displaced is lost despite the fact that one of these events has definitely occurred. Alice unambiguously distinguishes her measurement outcomes and Bob obtains at his disposal states of a single photon in superposition of two modes with controllable amplitude distortions. We consider the implementation of this protocol with a third participant who off-line prepares amplitude modulated unknown qubits and hand them to Alice for teleportation. In this interpretation, the success probability of the teleportation depends on the absolute value of the amplitude of the unknown qubit. In particular, highly unbalanced unknown qubits can be teleported with success probability close to one. We also consider some ways of demodulating amplitude modulated states by the receiver in order to increase the success probability of the teleportation of unknown qubits.

## 1. Introduction

Creation of a quantum computer capable of realizing algorithms such as quantum factoring [1] and quantum search [2] require both a design a universal set of gate operations for a large system and good fault-tolerant procedures to overcome inevitable imprecisions in unitary evolution of the physical system. There are many suggested approaches for quantum computers, but none of them are completely satisfactory, in the sense, that proposed methods are quite complex and can require an unacceptable number of additional operations to produce a specific desired reversible operations [3-5]. So, the question of resources (mechanisms, approaches, states) needed to realize scalable quantum computing is currently still open.

Light states are good candidates for quantum information processing [6]. So, single-qubit operations with photon states can be directly realized by linear optics methods. Immediate difficulties arise in the implementation of entangling gates with photon states as it is difficult to make photons interact with each other in a desired manner. The standard idea to implement entangling gates with optical states in practice is based on the teleportation protocol [7] and Bell-state measurement with linear optics [8]. The success probability of the Bell-state measurement with linear optics elements and photodetectors does not exceed 0.5 [8-10]. Controlled operations like to controlled $-X$ operation can be performed by simultaneous teleportation of two arbitrary qubits through entangled quantum channel [4], therefore success probability of such gates is limited to 0.25 [11-13]. Although, it is also worth mentioning the works about complete Bell state measurement realized by introduction of auxiliary photons [14] or making a joint detection of the polarization and angular momentum parity [15].

The other line of quantum information processing by optical qubits has been devoted to implementation with continuous variable states whose observable has a continuum of eigenvalues [16]. In the encoding, two coherent states $\left|-\alpha\right\rangle$ and $\left|\alpha\right\rangle$ are used as base elements. The Bell states measurement for coherent qubits can be performed in a nearly deterministic manner with $\alpha$ growing provided that single-qubit operations can be realized with the coherent states [17]. Entangled coherent states can be discriminated through photon number resolving detection (PNRD) [18] that is not easy to implement in practice. In general, approaches with discrete variable states can achieve fidelity close to unity but at the expense of the efficiency of processes (probabilistic restrictions), while continuous variable states suffer from strong sensitivity to loses and inevitable limited fidelities. Idea to combine the two approaches and use their best properties looks natural. The idea of hybridization between discrete variable and continuous variable states [19] can be exploited to have serious advantages in realization of quantum protocols and quantum computation [20-23]. Recently, some implementations of hybrid entanglement between a coherent qubit (superposition of coherent states (SCS)) and microscopic qubit of vacuum and single photon [24] and a single photon in polarization basis [25] were demonstrated.

Here, we develop novel way to implement quantum teleportation protocol. Hybrid entanglement is used for transmission of quantum information from superposition state of vacuum and single photon to the state of single photon occupying two modes. Driven force for the teleportation can be approximated by superposition of the displacement operators with opposite in sign amplitudes. The approach does not use Bell state formalism [26]. Coherent components of hybrid channel simultaneously displace unknown teleported qubit in indistinguishable manner on a highly transmissive beam splitter (HTBS) by the values that differ from each other only by sign $\pm\alpha$. Given operation is unconditional. Choice of the displacement amplitude $\alpha \ll 1$ greatly simplifies implementation of the protocol of quantum teleportation and removes the requirement of PNRD. Photons in teleported and coherent modes are measured and the teleported qubit disappear at the place of the measurement and it projects the single photon located arbitrary far from away into one of two possible states than can be unambiguously identified by the receiving party using 2 bits of classical information dispatched by the sender. Bob's two states undergo amplitude distortion. We consider the implementation with the help of a third participant whose actions on the amplitude modulation of the original unknown qubit are autonomously performed and are not included in the protocol. Amplitude demodulation of an unknown qubit by the receiver to increase his success probability to recover original qubit is also considered. Mathematical apparatus for the development of the protocol is based on representation of the displaced number states in terms of the number states [23,27]. This method has been used to generate even and odd SCS states of large amplitude by subtraction of photons from squeezed coherent state regardless of the number of subtracted photons [28,29] as well as to consider feasibility of one-dimensional rotations of coherent states (Hadamard gate) [30].

A detailed review of the displacement operations [31,32] is presented in section 2. In section 3, we describe direct implementation of the quantum teleportation protocol. We show its feasibility and discuss the problem of amplitude distortion of obtained qubits. In section 4, we discuss the methods of increase of its efficiency by controllable amplitude modulation of original unknown qubit performed by third participant. In section 5, we discuss methods of demodulation of unknown qubit. Section 6 generalizes key moments of the studied quantum teleportation protocol. Additional auxiliary mathematical apparatus with displaced states of photons is presented in Appendix A. Appendix B concerns applicability of the mechanism to generation of the hybrid state and procedure of amplitude demodulation performed by strong coherent pump.

## 2 Realization of displacement operators

Before considering the protocol of quantum teleportation of unknown qubit through a hybrid channel, let us consider interaction of strong coherent field with arbitrary state on HTBS

$$BS = \begin{vmatrix} t & -r \\ r & t \end{vmatrix},\tag{1}$$

where $t$, $r$ are the transmittance $t \to 1$ and reflectance $r \to 0$, respectively, satisfying the normalization condition $t^2 + r^2 = 1$. We consider that the parameters $t$ and $r$ are the real values. So, interaction of two coherent states $|\alpha\rangle_1 |\beta\rangle_2$ with amplitudes $\alpha$ and $\beta$, respectively, gives

$$BS|\alpha\rangle_1 |\beta\rangle_2 = |\alpha t + \beta r\rangle_1 |\beta t - \alpha r\rangle_2.\tag{2}$$

where subscripts denote the state modes. Let $\rho_1$ be an arbitrary state which can be written in terms of the Glauber$-$Sudarshan $P-$function as

$$\rho_1 = \int d^2\alpha P(\alpha)|\alpha\rangle_1\langle\alpha|. \qquad (3)$$

Consider interaction of the state with coherent state on HTBS (1). By virtue of (2), we have

$$BS(\rho_1 \otimes |\beta\rangle_2\langle\beta|)BS^+ = \int d^2\alpha P|\alpha t + \beta r\rangle_1\langle\alpha t + \beta r| \otimes |\beta t - \alpha r\rangle_2\langle\beta t - \alpha r|. \qquad (4)$$

where the notation $\otimes$ means tensor product. The integral (4) can be transformed to

$$\int d^2\alpha P|\alpha + \gamma\rangle_1\langle\alpha + \gamma| \otimes |\beta\rangle_2\langle\beta| = \int d^2\alpha PD(\gamma)|\alpha\rangle_1\langle\alpha|D^+(\gamma) \otimes |\beta\rangle_2\langle\beta|. \qquad (5)$$

in the limit case of [33]

$$t \to 1, \ r \to 0, \text{ but } \beta r \to \gamma, \qquad (6)$$

where the displacement operator (A1) with $\gamma$ being the displacement amplitude is used. Then finally, we can rewrite (4)

$$BS(\rho_1 \otimes |\beta\rangle_2\langle\beta|)BS^+ \approx D_1(\gamma)\rho_1 D_1^+(\gamma) \otimes |\beta\rangle_2\langle\beta|, \qquad (7)$$

which in the case of pure state $\rho_1 = |\Psi\rangle_1\langle\Psi|$ implies

$$BS(|\Psi\rangle_1 \otimes |\beta\rangle_2) \approx D_1(\gamma)|\Psi\rangle_1 \otimes |\beta\rangle_2. \qquad (8)$$

The condition (6) means that amplitude of auxiliary coherent state $|\beta\rangle_2$ tends to infinity $\beta \to \infty$ [33]. In real experiment, the amplitude of the coherent state to implement deterministic displacement of arbitrary state must be chosen to be sufficiently large but not infinite. If we apply the coherent state with negative amplitude $|-\beta\rangle$, then the result of interaction on HTBS can be approximated by

$$BS(|\Psi\rangle_1 \otimes |-\beta\rangle_2) \approx D_1(-\gamma)|\Psi\rangle_1 \otimes |-\beta\rangle_2. \qquad (9)$$

Expressions (8,9) are applicable to consideration of the teleportation protocol shown in Fig. 1. Suppose Alice wants to teleport unknown qubit

$$|\varphi\rangle_2 = a_0|0\rangle_2 + a_1|1\rangle_2, \qquad (10)$$

to Bob located at a considerable distance apart from Alice. Alice cannot send this qubit directly but she has at her disposal a part of quantum channel,

$$|\Psi\rangle_{134} = (|0,-\beta\rangle_1|01\rangle_{34} + |0,\beta\rangle_1|10\rangle_{34})/\sqrt{2}, \qquad (11)$$

which is created in advance and connects Alice and Bob. Here, the notation for displaced number states (A2) is used. The real amplitude $\beta$ is assumed to be positive $\beta > 0$ throughout the consideration. The quantum channel is an entangled hybrid state, which consists of coherent components belonging to Alice (mode 1 ) and one photon (dual-rail photon), which simultaneously take two modes (modes 3 and 4 ) at Bob's disposal. We consider generation of the hybrid entangled state in Appendix B. Note also the fact the teleported qubit (10) is defined in the basis $\{|0\rangle, |1\rangle\}$, while Bob's photon is determined in base $\{|01\rangle, |10\rangle\}$. Alice mixes unknown qubit (10) with her coherent components on HTBS as shown in Fig. 1. Result of the mixing is given by

$$BS_{12}(|\Psi\rangle_{134}|\varphi\rangle_2) = (BS_{12}(|0,-\beta\rangle_1|\varphi\rangle_2)01\rangle_{34} + BS_{12}(|0,\beta\rangle_1|\varphi\rangle_2)10\rangle_{34})/\sqrt{2}, \qquad (12)$$

due to linearity of the unitary beam splitter operator (1). Let us consider the action of coherent components on the teleported state separately. So, output state being result of mixing of the coherent state $|0,-\beta\rangle$ with the unknown qubit (10) on HTBS is written

$$BS_{12}\big(|0,-\beta\rangle_1|\varphi\rangle_2\big)|01\rangle_{34} = BS_{12}(D_1(\beta)D_2(-\alpha))BS_{12}^+ BS_{12}\big(|0\rangle_1 D_2(\alpha)|\varphi\rangle_2\big)|01\rangle_{34} =$$

$$D_1(-\beta/t)D_2(0)BS_{12}\big(|0\rangle_1\big(a_0|0,\alpha\rangle_2 + a_1|1,\alpha\rangle_2\big)\big)|01\rangle_{34} =$$

$$FD_1(-\beta/t)BS_{12}\left(|0\rangle_1\sum_{m=0}^{\infty}\big(a_0 c_{0m}(\alpha)+a_1 c_{1m}(\alpha)\big)|m\rangle_2\right)|01\rangle_{34} =$$ . (13)

$$F|0,-\beta/t\rangle_1\left(\sum_{m=0}^{\infty}t^m\big(a_0 c_{0m}(\alpha)+a_1 c_{1m}(\alpha)\big)|m\rangle_2\right)|01\rangle_{34} +$$

$$FD_1(-\beta/t)\left(\sum_{m=0}^{\infty}\big(a_0 c_{0m}(\alpha)+a_1 c_{1m}(\alpha)\big)|X^{(m)}\rangle_{12}\right)|01\rangle_{34}$$

Using the same calculation technique, one obtains output state being result of interaction of the coherent state $|0,\beta\rangle$ and the teleported state (10) as inputs to the HTBS in Fig. 1

$$BS_{12}\big(|0,\beta\rangle_1|\varphi\rangle_2\big)|10\rangle_{34} = BS_{12}(D_1(\beta)D_2(\alpha))BS_{12}^+ BS_{12}\big(|0\rangle_1 D_2(-\alpha)|\varphi\rangle_2\big)|10\rangle_{34} =$$

$$D_1(\beta/t)D_2(0)BS_{12}\big(|0\rangle_1\big(a_0|0,-\alpha\rangle_2 + a_1|1,-\alpha\rangle_2\big)\big)|10\rangle_{34} =$$

$$FD_1(\beta/t)BS_{12}\left(|0\rangle_1\sum_{m=0}^{\infty}\big(a_0 c_{0m}(-\alpha)+a_1 c_{1m}(-\alpha)\big)|m\rangle_2\right)|10\rangle_{34} =$$ . (14)

$$F|0,\beta/t\rangle_1\left(\sum_{m=0}^{\infty}(-1)^m t^m\big(a_0 c_{0m}(\alpha)-a_1 c_{1m}(\alpha)\big)|m\rangle_2\right)|10\rangle_{34} +$$

$$FD_1(\beta/t)\left(\sum_{m=0}^{\infty}(-1)^m\big(a_0 c_{0m}(\alpha)-a_1 c_{1m}(\alpha)\big)|X^{(m)}\rangle_{12}\right)|10\rangle_{34}$$

Here, we made use of unitary properties of the beam splitter and displacement operators $BS(BS)^+ = I$ and $D(\alpha)(D(\alpha))^+ = D(\alpha)D(-\alpha) = I$, respectively, where $I$ is an identity operator and notation + means Hermitian conjugate [34]. The total factor $F$ is introduced in Appendix A after formula (A5). The displacement amplitude $\alpha$ is chosen to fulfill $\alpha = \beta r/t$. Since the amplitude of the coherent states is chosen to be a positive value $\beta > 0$, then the displacement amplitude takes the positive values $\alpha > 0$. The decompositions of the coherent and displaced single photon states [23] over the number states (A9,A10) are used in derivation of (13,14). The matrix elements (A7) are written as a product of $\alpha$ in some power and another factor bracketed. The bracketed factor is a certain polynomial in absolute values of the displacement amplitude and does not affect the shape of the matrix element in the case when we rotate the displacement amplitudes in the phase plane by some angle. Conversely, the first factor determines the phase of the matrix element when the displacement amplitude is rotated in the phase plane. It is key moment to learning discrete-continuous interaction on HTBS. So, the matrix elements of the zero row $c_{0n}(\alpha)$ (A7) change as $(-1)^n$ under the change of the displacement amplitude on opposite $\alpha \to -\alpha$ (A16). They are elements of the coherent state. At the same time, the elements of the first row (they are elements of the displaced single photon) change as $(-1)^{n-1}$ under the change of the displacement amplitude on opposite (A17). This difference in the behavior of the matrix elements is akin to a nonlinear action of two-qubit gate controlled-Z gate. The change in the sign of the matrix elements under change of the displacement amplitude on opposite (A13-A17) allows us to present result of mixing (12) as entangled state. Expressions (13,14) involve output state being result of passing the number state $|m\rangle$ through the HTBS [34]

$$BS\big(|0\rangle_1|m\rangle_2\big) = \big((ra_1^+ + ta_2^+)^m/\sqrt{m!}\big)|00\rangle_{12} = |X^{(m)}\rangle_{12} + t^m|0\rangle_1|m\rangle_2,$$ (15)

where

$$|X^{(m)}\rangle_{12} = \big(1/\sqrt{m!}\big)\sum_{k=0}^{m-1}\frac{r^{m-k}t^k m!}{k!(m-k)!}\sqrt{k!(m-k)!}|m-k\rangle_1|k\rangle_2 .$$ (16)

Summing up the formulas (13,14), one obtains the final state

$$BS\big(|\Psi\rangle_{1234}|\varphi\rangle_2\big) = |\Delta_1\rangle_{1234} + |\Delta_2\rangle_{1234},\qquad(17)$$

where

$$|\Delta_1\rangle_{1234} = \frac{F}{\sqrt{2}}\begin{pmatrix}|0,-\beta/t\rangle_1\left(\sum_{m=0}^{\infty}t^m\big(a_0 c_{0m}(\alpha)+a_1 c_{1m}(\alpha)\big)|m\rangle_2\right)|01\rangle_{34} + \\ |0,\beta/t\rangle_1\left(\sum_{m=0}^{\infty}(-1)^m t^m\big(a_0 c_{0m}(\alpha)-a_1 c_{1m}(\alpha)\big)|m\rangle_2\right)|10\rangle_{34}\end{pmatrix},\qquad(18)$$

$$|\Delta_2\rangle_{1234} = \frac{F}{\sqrt{2}}\begin{pmatrix}D_1(-\beta/t)\left(\sum_{m=0}^{\infty}\big(a_0 c_{0m}(\alpha)+a_1 c_{1m}(\alpha)\big)|X^{(m)}\rangle_{12}\right)|01\rangle_{34} + \\ D_1(\beta/t)\left(\sum_{m=0}^{\infty}(-1)^m\big(a_0 c_{0m}(\alpha)-a_1 c_{1m}(\alpha)\big)|X^{(m)}\rangle_{12}\right)|10\rangle_{34}\end{pmatrix}.\qquad(19)$$

The contribution of first term (18) in sum (17) prevails over the second (19) in the case of $t\to1$, $r\to0$. If $t\to1$ and $r\to0$, then the final state (17) tends to

$$BS_{12}\big(|\Psi\rangle_{134}|\varphi\rangle_2\big)\to|\Delta_1^{(id)}\rangle_{1234},\qquad(20)$$

where the ideal normalized state becomes

$$|\Delta_1^{(id)}\rangle_{1234} = \frac{F}{\sqrt{2}}\begin{pmatrix}|0,-\beta\rangle_1\left(\sum_{m=0}^{\infty}\big(a_0 c_{0m}(\alpha)+a_1 c_{1m}(\alpha)\big)|m\rangle_2\right)|01\rangle_{34} + \\ |0,\beta\rangle_1\left(\sum_{m=0}^{\infty}(-1)^m\big(a_0 c_{0m}(\alpha)-a_1 c_{1m}(\alpha)\big)|m\rangle_2\right)|10\rangle_{34}\end{pmatrix}.\qquad(21)$$

where now amplitude $\beta$ of the quantum channel (11) and the displacement amplitude $\alpha$ are connected by

$$\alpha = \beta\sqrt{1-t^2}.\qquad(22)$$

In real case of HTBS with non-zero reflectance, the state (21) can only approximate the real state. By analogy with (8,9), we can present final state as

$$BS_{12}\big(|\Psi\rangle_{134}|\varphi\rangle_2\big)\approx|\Delta_1^{(id)}\rangle_{1234}.\qquad(23)$$

The fidelity of such approximation can be evaluated by

$$Fid = \left|\langle\Delta_1^{(id)}|\big(|\Delta_1\rangle+|\Delta_2\rangle\big)\right|^2.\qquad(24)$$

Unit fidelity means the states are identical to each other [26]. Substituting the considered states into (24), one obtains analytical expression for the fidelity

$$Fid = F^4\frac{\exp\big(-|\beta|^2(1-1/t)^2\big)}{4}\left(\sum_{m=0}^{\infty}t^m\big|f_m(-\alpha)\big|^2 + \sum_{m=0}^{\infty}t^m\big|f_m(\alpha)\big|^2\right)^2,\qquad(25)$$

where

$$f_m(\pm\alpha) = a_0 c_{0m}(\pm\alpha)+a_1 c_{1m}(\pm\alpha).\qquad(26)$$

Here, we neglect the terms proportional to $\sim r^2$ whose contribution is substantially insignificant in the case of $r\ll1$. The fidelity becomes $Fid=1$ in the case of $t=1$ due to normalization conditions (A20). Now, we can talk about approximation of the interaction of the entangled state (11) with the teleported state (10) by the following operator

$$\Omega = \big(|0,-\beta\rangle_1|01\rangle_{34}D_2(\alpha)+|0,\beta\rangle_1|10\rangle_{34}D_2(-\alpha)\big)\big/\sqrt{2},\qquad(27)$$

which is applied to the teleported qubit (10) to give

$$\Omega|\varphi\rangle_2 = |\Delta_1^{(id)}\rangle_{1234}.\qquad(28)$$

Operator $\Omega$ is a superposition of the terms composed of the states and displacement operators with absolute equal but opposite in sign amplitudes. The state (21) is used in analysis of the quantum teleportation protocol of unknown qubit. Note the operation (27) can be realized in experimental scenario using the same technique as in [31,32]. For example, if we pass the SCS through balanced beam splitter, we obtain entangled superposition of coherent states

$$N_-\left(\left|0,-\sqrt{2}\beta\right\rangle_1 - \left|0,\sqrt{2}\beta\right\rangle_1\right) \to N_-\left(\left|0,-\beta\right\rangle_1 \left|0,-\beta\right\rangle_2 - \left|0,\beta\right\rangle_1 \left|0,\beta\right\rangle_2\right),$$

which can be represented as the following operator

$$N_-\left(\left|0,-\beta\right\rangle_1 D_2(-\beta) - \left|0,\beta\right\rangle_1 D_2(\beta)\right)\left|0\right\rangle_2,$$

acting on vacuum. Comparing the formula and formulas (27,28) , we can see their conformity.

In general case of $t \neq 1$, the fidelities (25) are the functions of the amplitudes of the teleported state, its phase relationships, displacement amplitude $\alpha$ due to presence of the expansion coefficients $c_{nm}(\alpha)$ in expression for the fidelities as well as transmittance $t$. Numerical calculations show that the fidelities though depend on the parameters of the teleported state, this relationship is not significant. The fidelity is largely determined by two parameters $\alpha$ and $t$. Corresponding plot in figures 2 shows the fidelity (25) as function of $\alpha$ and $t$ for equal modulo amplitudes of the teleported qubit $a_0 = a_1 = 1/\sqrt{2}$. Similar dependencies little different from each other are observed for other amplitudes values. Realization of the displacement operator with help of HTBS is extremely sensitive to the parameters $\alpha$ and $t$. The fidelities take unit value in the case of $t = 1$ regardless the displacement amplitude $\alpha$ [33]. But they fairly quickly fall almost to zero with increasing displacement parameter $\alpha$ and decreasing transmittance $t$. In order to achieve high fidelity $Fid(\alpha) > 0.99$ it is required to choose a beam splitter with extremely high transmittance $t \cong 1$ or to consider displacement of the initial state on relatively small value $\alpha < 1$. Thus, the efficiency of the displacement method with HTBS cannot be recognized to be high as the operation can only be effectively performed on low values of the displacement amplitudes $|\alpha| \leq 1$ for those values of transmittance $t$, which could be used in practice. We can hardly say that the generated entangled hybrid state (11) involve macroscopic (e. g., visible by eye [31,32]) states. Displaced entangled states of low amplitude were also used for the implementation of the dense coding protocol in [35].

## 3 Direct implementation of quantum teleportation protocol

In the previous section, we considered deterministic interaction of the quantum channel (11) with the teleported unknown qubit (10) on HTBS. As result of the interaction, the teleported qubit is displaced on values $\pm\alpha$ in such a way that all information on what quantity the teleported qubit has been displaced disappears. This uncertainty underlies the successful implementation of the protocol. Ideal output state (21) can be rewritten in the terms of even/odd SCS as

$$BS\left(\left|\Psi\right\rangle_{134}\left|\varphi\right\rangle_2\right) = \sum_{n=0}^{\infty} g_n \left( \begin{array}{c} \dfrac{1}{N_+}\left|even\right\rangle_1 \dfrac{N_n}{\sqrt{2}}\left(\left(a_0 + A_n a_1\right)\left|01\right\rangle_{34} + (-1)^n \left(a_0 - A_n a_1\right)\left|10\right\rangle_{34}\right) + \\ \dfrac{1}{N_-}\left|odd\right\rangle_1 \dfrac{N_n}{\sqrt{2}}\left(\left(a_0 + A_n a_1\right)\left|01\right\rangle_{34} - (-1)^n \left(a_0 - A_n a_1\right)\left|10\right\rangle_{34}\right) \end{array} \right) \left|n\right\rangle_2 , \quad (29)$$

where the following parameters are introduced

$$A_n = \frac{n - |\alpha|^2}{\alpha} , \tag{30}$$

$$N_n = \left(|a_0|^2 + |A_n|^2 |a_1|^2\right)^{-0.5} = \left(1 + \left(|A_n|^2 - 1\right)|a_1|^2\right)^{-0.5} , \tag{31}$$

$$g_n = F \frac{\alpha^n}{2N_n \sqrt{n!}} . \tag{32}$$

$$\left|even\right\rangle_1 = N_+\left(\left|0,-\beta\right\rangle + \left|0,\beta\right\rangle\right)_1 , \tag{33a}$$

$$\left|odd\right\rangle_1 = N_-\left(\left|0,-\beta\right\rangle - \left|0,\beta\right\rangle\right)_1 , \tag{33b}$$

where the factors $N_\pm = \left(2\left(1 \pm \exp\left(-2|\beta|^2\right)\right)\right)^{-1/2}$ are the normalization parameters of the even/odd SCS.

After that, Alice performs two types of measurements: parity measurement in first mode and photon number measurement in the second mode. Having performed the measurements, Alice generates the following states at Bob's disposal either

$$\frac{N_n}{\sqrt{2}}\left(\left(a_0 + A_n a_1\right)|01\rangle_{34} + \left(a_0 - A_n a_1\right)|10\rangle_{34}\right), \text{if } \left(j=0, n=2m\right) \text{ and } \left(j=1, n=2m+1\right), \quad (34a)$$

or

$$\frac{N_n}{\sqrt{2}}\left(\left(a_0 + A_n a_1\right)|01\rangle_{34} - \left(a_0 - A_n a_1\right)|10\rangle_{34}\right), \text{if } \left(j=0, n=2m+1\right) \text{ and } \left(j=1, n=2m\right), \quad (34b)$$

where the notation $\left(j=0/j=1, n=2m/2m+1\right)$ indicates that *even/odd* number of photons is registered in the first mode and $n = 2m$ or $n = 2m+1$ photons is fixed in the second mode by Alice. Value $j=0$ corresponds to even SCS and $j=1$ encodes odd SCS. For example, record $\left(j=0, n=2m\right)$ in (34a) implies that Alice has detected even SCS by measuring the *coherent* mode and $n$ photons in the second mode where $n = 2m$ is even number. Alice can record the results of her measurements by a string of two digits $\left(j, n\right)$. The first number $j$ is the binary one and displays the result of her parity measurement, while the second number $n$ is a decimal and displays the number of measured photons in the teleported qubit. A decimal number $n$ must be translated into binary code before sending the message to Bob increasing total bit line length. Bob reads message and realizes the Pauli $Z-$operation $Z^{j+par(n)}$ on his dual-rail single photon by phase shift in one of two modes by $\pi$, where $j = 0,1$ and $par(n)$ means the parity of number $n$. After this, Bob performs Hadamard operation $H$ on his dual-rail single photon regardless of Alice's measurement outcomes

$$HZ^{j+par(n)}\frac{N_n}{\sqrt{2}}\begin{vmatrix} a_0 + A_n a_1 \\ (-1)^j\left(a_0 - A_n a_1\right) \end{vmatrix} = |\psi_n\rangle = N_n\begin{vmatrix} a_0 \\ A_n a_1 \end{vmatrix}, \quad (35)$$

where $H$ is Hadamard transformation

$$H = \frac{1}{\sqrt{2}}\begin{vmatrix} 1 & 1 \\ 1 & -1 \end{vmatrix}, \quad (36a)$$

and $Z$ matrix is

$$Z = \begin{vmatrix} 1 & 0 \\ 0 & -1 \end{vmatrix}. \quad (36b)$$

The Hadamard operation on a single photon can be implemented using the balanced beam splitter and phase shift operations [36]. The states $|\psi_n\rangle$ (Eq. (35)) is written in dual-rail basis of single photon $\left\{|01\rangle, |10\rangle\right\}$ unlike initial $\left\{|0\rangle, |1\rangle\right\}$. The states are not original (10) as they involve additional factors $N_n$ and $N_n A_n$, respectively. But the states $|\psi_n\rangle$ conserve phase relations with original one. Since the states include additional amplitude factors known to Alice and not affecting the phase relations, then such states can be called amplitude modulated (AM). Subscript $n$ corresponds to number of photons measured in the teleported state and defines parameter of amplitude modulation (30) of the output state. The reverse process can be called the demodulation of the AM states.

The success probability to generate the states (35) at Bob's station depends on parameters of the teleported state, namely, on the absolute values of the amplitudes either $|a_0|$ or $|a_1|$

$$P_n(\alpha) = \exp\left(-|\alpha|^2\right)\frac{|\alpha|^{2n}\left(1 + \left(|A_n|^2 - 1\right)|a_1|^2\right)}{n!}. \quad (37)$$

One can directly show using (A20) that the total probability of the events is equal to one $\sum_{n=0}^{\infty} P_n(\alpha) = 1$. Note that the additional amplitude factors and, as consequence, the dependence of the success probability on the absolute values of the teleported qubit mathematically arise as a result of the fact that the coherent state and displaced single photon state are transformed differently when projecting the states onto measurement basis of the number states. Corresponding three-dimensional plots $P_n$ for $n = 0,1,2,3$ as functions of $\alpha$ and $|a_1|$ are presented in Fig. 3(a-d). Variation range of the displacement amplitude is chosen within $\alpha \in [-0.5, 0.5]$. Plots in figure 3(a-d) show that the probabilities $P_0$ and $P_1$ prevail over other probabilities (especially over the probabilities $P_n$ of higher order with $n > 3$) in a wide range of change of the displacement amplitude $\alpha$ and absolute values of the qubit amplitude $|a_1|$. These plots allow us to claim the less we choose the value of the displacement amplitude, the greater we observe the preponderance of $P_0$ and $P_1$ over the other probabilities $P_n$ with $n > 1$ $(P_0, P_1 \gg P_n)$. The plot in figure 3(e) made for $\alpha = 0.03$ fully supports this conclusion. In the case, the sum probability takes a minimum value $P_0 + P_1 \geq P_{\min} = 0.9982$. It is worth noting that the probabilities $P_0$ and $P_1$ evolve in opposite relation to each other, that is, if $P_0$ falls, then $P_1$ increases, and vice versa. The probabilities of higher orders $P_n$ with $n > 1$ take much smaller values in the entire range of variation of the absolute amplitude $|a_1|$.

It is unlikely that such implementation of the quantum teleportation protocol of unknown qubit can be practical since it requires the use of special detectors capable to determine the parity of even/odd SCS and discriminate among incoming photons. In addition, an increase in the measurement outcomes leads to an increase of the classical information flow from Alice to Bob which can be hardly considered an advantage of the approach. Therefore, reducing the displacement amplitude $\alpha$ becomes better strategy in terms of implementation of the studied protocol in practice. Moreover, the amplitude of the coherent components of the hybrid channel (11) can be significantly reduced. For example, consider the beam splitter (1) with transmission coefficient $T = t^2 = 0.99$ which in combination with the small displacement amplitude value $\alpha = 0.03$ gives small value $\beta = 0.3$ (Eq. (22)) of amplitude of the coherent components. Use of the SCS with the small amplitude value allows us to approximate them by the following number states with high fidelity

$$|even\rangle \approx |0\rangle, \tag{38a}$$

$$|odd\rangle \approx |1\rangle. \tag{38b}$$

Indeed, the probability distributions of the *even/odd* SCS, for example, with $\beta = 0.3$

$$P_{2n}^{even}(\beta) = 4N_+^2(\beta)\exp\left(-|\beta|^2\right)\left(|\beta|^2\right)^{2n}/(2n)!, \tag{39a}$$

$$P_{2n+1}^{odd}(\beta) = 4N_-^2(\beta)\exp\left(-|\beta|^2\right)\left(|\beta|^2\right)^{2n+1}/(2n+1)!, \tag{39b}$$

take the following values

$P_0^{even}(\beta = 0.3) = 0.996$, $\quad P_2^{even}(\beta = 0.3) = 0.004$, $\quad P_4^{even}(\beta = 0.3) = 2.72 \times 10^{-6}$,

$P_1^{odd}(\beta = 0.3) = 0.9986$, $\quad P_3^{odd}(\beta = 0.3) = 0.0013$, $\quad P_5^{even}(\beta = 0.3) = 4.9 \times 10^{-8}$.

Probabilities $P_0^{even}(\beta = 0.3)$ and $P_0^{odd}(\beta = 0.3)$ prevail over other ones that enables to make use of approximation (38). This is more than enough to take advantage of commercially achievable avalanche photodiode (APD) being a highly sensitive semiconductor electronic device that exploits the photoelectric effect to convert light to electricity and that can ideally operate in on-off regime $|0\rangle\langle 0| + \sum_{n=1}^{\infty} |n\rangle\langle n|$. Thus, the parity measurement in the case of $\beta < 1$ can be replaced by APD able to distinguish outcomes from vacuum and single photon. Use of the small amplitudes $\beta < 1$ of the quantum channel (11) guaranties performance of $\alpha << 1$ (22). Then, registration of two photons (not mentioning light pulses with a larger number of photons) in the teleported mode is unlikely as the probability of such events is less than one percent (Figs. 3(a-d)). This means that APD can also be used in the teleported mode in the case of $\alpha << 1$.

Finally, the quantum teleportation protocol of unknown qubit can be described in simpler form in the case of $\alpha << 1$ instead of (29)

$$BS\left(|\Psi\rangle_{134}|\varphi\rangle_2\right) = g_0 \begin{pmatrix} \dfrac{1}{N_+}|00\rangle_{12} \dfrac{N_0}{\sqrt{2}}\left(\left(a_0 + A_0 a_1\right)|01\rangle_{34} + \left(a_0 - A_0 a_1\right)|10\rangle_{34}\right) + \\ \dfrac{1}{N_-}|10\rangle_{12} \dfrac{N_0}{\sqrt{2}}\left(\left(a_0 + A_0 a_1\right)|01\rangle_{34} - \left(a_0 - A_0 a_1\right)|10\rangle_{34}\right) \end{pmatrix} + $$
$$g_1 \begin{pmatrix} \dfrac{1}{N_+}|01\rangle_{12} \dfrac{N_1}{\sqrt{2}}\left(\left(a_0 + A_1 a_1\right)|01\rangle_{34} - \left(a_0 - A_1 a_1\right)|10\rangle_{34}\right) + \\ \dfrac{1}{N_-}|11\rangle_{12} \dfrac{N_1}{\sqrt{2}}\left(\left(a_0 + A_1 a_1\right)|01\rangle_{34} + \left(a_0 - A_1 a_1\right)|10\rangle_{34}\right) \end{pmatrix}, \quad (40)$$

with fidelity prevailing $> 0.99$. Subsequent Alice's measurement in the base $\{|00\rangle, |01\rangle, |10\rangle, |11\rangle\}$ instead of $\{|even/odd\rangle, |n\rangle\}$ allows Bob to receive one of two possible states either $|\psi_0\rangle$ or $|\psi_1\rangle$ (Eq. (35)). Note only that Alice's measured outcomes are the same as if she had performed Bell-state measurement $|00\rangle_{12}, |10\rangle_{12}, |01\rangle_{12}$, and $|11\rangle_{12}$, respectively [7,26]. Bob does not know exactly which state he has at his disposal and he needs Alice to help him to identify them. Alice encodes her measurement outcomes by two bits in length as in [7] in full compliance with her measured results $|jk\rangle_{12} \rightarrow (j,k)$, where $j,k = 0,1$ (Fig. 1). Bob reads the bits and decides whether or not to apply the $Z$ operation $\left(Z^{j+k}\right)$ to his single photon with subsequent application of the Hadamard operation. The second bit is assigned for Bob to unambiguously determine which state either $|\psi_0\rangle$ $(k = 0)$ or $|\psi_1\rangle$ $(k = 1)$ he obtained. Summarizing all of the above, the results of this section are reflected in Table 1 for the case of $\alpha << 1$. Note only Bob obtain the states in base $\{|01\rangle, |10\rangle\}$ instead of original one (10) defined in two-dimensional Hilbert space with base elements $\{|0\rangle, |1\rangle\}$.

| Measurement outcomes | Obtained states | Success probabilities |
|---|---|---|
| (0,0),  (1,0) | $|\psi_0\rangle$   (AM) | $P_0(\alpha)$ |
| (0,1),  (1,1) | $|\psi_1\rangle$   (AM) | $P_1(\alpha)$ |

**Table 1**. Concise generalization of quantum teleportation protocol with hybrid entangled state (11) realized by Alice directly. AM means amplitude modulated state. Sum of the success probabilities is almost equal to one $P_0(\alpha) + P_1(\alpha) \approx 1$ in the case of $\alpha << 1$.

## 4 Quantum teleportation of amplitude modulated unknown qubit

We have shown the use of small values of the displacement amplitude $\alpha$ on which we need to displace the teleported qubit allows us significantly to increase effectiveness of the protocol. Parity measurement and photon number resolving measurement can be replaced by on-off measurement that greatly increases the chances to implement the protocol in practice. Hybrid quantum state (11) with a sufficiently small value of the amplitude of the coherent states is also an advantage of the protocol. Nevertheless, the problem of amplitude demodulation of the output qubits remains. To increase the efficiency of the protocol we are going to consider quantum teleportation of initially AM unknown qubits. This version of the protocol can be considered in one of the possible strategies for its development with a third party (assume Charles). Charles's actions are separately singled out so that his probabilistic methods of obtaining AM qubits do not affect the total success probability. Indeed, Charles's actions to prepare AM qubits can be attributed to preliminary as well as the realization of hybrid entangled state (11). The difference between the preparation of an unknown qubit and AM unknown qubit is that the AM qubit is subjected to a controlled change in amplitudes, that is, it must be known in advance that such a modification under qubit has been made. If there is an opportunity to generate AM qubits in advance, then it is possible to do without Charles for such implementation of the protocol. Assume that Charles prepares the AM unknown qubits for Alice from original unknown qubit and transmits them to Alice. In the interpretation, Charles's actions are considered preparatory and are made off-line.

Suppose that Charles can prepare for Alice the following AM qubit from initial original one (10)

$$\left| \varphi_0^{(in)} \right\rangle_2 = N_0^{(in)} \begin{vmatrix} a_0 \\ A_0^{-1} a_1 \end{vmatrix}, \tag{41}$$

where amplitudes $a_0$ and $a_1$ are unknown parameters, $A_0$ is a modulation factor given by (30) and the normalization factor $N_0^{(in)}$ is given by $N_0^{(in)} = \left( |a_0|^2 + |A_0|^{-2} |a_1|^2 \right)^{-0.5} = \left( 1 + \left( |A_0|^{-2} - 1 \right) |a_1|^2 \right)^{-0.5}$. The protocol is also performed as described in Section 2 and the results are shown in Table 2 in the case of $\alpha \ll 1$.

| Measurement outcomes | Obtained states | Success probabilities |
|---|---|---|
| $(0,0)$, $(1,0)$ | $\left| \psi_{00}^{(out)} \right\rangle$ (original) | $P_{00}(\alpha)$ |
| $(0,1)$, $(1,1)$ | $\left| \psi_{10}^{(out)} \right\rangle$ (AM) | $P_{10}(\alpha)$ |

**Table 2**. Results of quantum teleportation protocol of initial AM unknown qubit (41) in the case of $\alpha \ll 1$ when preparatory actions of Charles to prepare initial AM qubit from original unknown one are not involved in the protocol. Bob can directly restore the original qubit with probability $P_{00}(\alpha)$ and the following relation $P_{00}(\alpha) + P_{10}(\alpha) \approx 1$ is performed in the case.

Here, the following notations are introduced

$$\left| \psi_{00}^{(out)} \right\rangle = \begin{vmatrix} a_0 \\ a_1 \end{vmatrix}, \tag{42}$$

$$\left| \psi_{n0}^{(out)} \right\rangle = N_{n0} \begin{vmatrix} a_0 \\ A_n A_0^{-1} a_1 \end{vmatrix}, \tag{43}$$

where the normalization factor is $N_{n0} = \left( |a_0|^2 + |A_n|^2 |A_0|^{-2} |a_1|^2 \right)^{-0.5} = \left( 1 + \left( |A_n|^2 |A_0|^{-2} - 1 \right) |a_1|^2 \right)^{-0.5}$. Corresponding success probabilities are given by

$$P_{00}(\alpha) = \exp\left( -|\alpha|^2 \right) N_0^{(in)2} = \frac{\exp\left( -|\alpha|^2 \right)}{1 + \left( |A_0|^{-2} - 1 \right) |a_1|^2}, \tag{44}$$

$$P_{n0}(\alpha) = \exp\left( -|\alpha|^2 \right) \frac{|\alpha|^{2n}}{n!} \frac{N_0^{(in)2}}{N_{n0}^2} = \exp\left( -|\alpha|^2 \right) \frac{|\alpha|^{2n}}{n!} \frac{1 + \left( |A_n|^2 |A_0|^{-2} - 1 \right) |a_1|^2}{1 + \left( |A_0|^{-2} - 1 \right) |a_1|^2}, \tag{45}$$

where $n = 1$ is used in Table 2.

The distribution (44,45) differs one (37). It is possible directly to show the distribution (44,45) is normalized $\sum_{n=0}^{\infty} P_{n0}(\alpha) = 1$. Plots of the probability dependencies $P_{00}$, $P_{10}$ and $P_{20}$ on $|a_1|$ are shown in figures 4(a-d) for different values of the displacement parameter $\alpha$. As can be seen from the plots, there is a range of values of $|a_1|$, at which the success probability of teleportation is more of 0.5. Participants of the teleportation protocol (Alice and Bob) may be fortunate even not suspecting about it and teleport unknown qubit (10) with success probability close to one if highly unbalanced qubit with $|a_0| << |a_1|$ is used. The highly unbalanced qubits can be interpreted as located near the poles of the Bloch sphere. Increase of the displacement parameter $\alpha$ enables to increase the range of values $|a_1|$ for which the success probability of the teleportation is more than 0.5 (Figs. 4(c,d)). But the increase of the displacement amplitude $\alpha$ is restricted from a practical point of view. Contribution of the state $|\psi_{20}^{(out)}\rangle$ (curve 3 in Fig. 4(c,d)) may become essential in the case of increase of the displacement amplitude $\alpha$.

Consider the case when Charles can produce another type of amplitude modulation of the unknown qubit (10)

$$|\varphi_1^{(in)}\rangle_2 = N_1^{(in)} \left| \begin{array}{c} a_0 \\ A_1^{-1}a_1 \end{array} \right|, \qquad (46)$$

where quantity $A_1$ is the modulation factor given by (30) and the normalization factor $N_1^{(in)}$ is $N_1^{(in)} = \left( |a_0|^2 + |A_1|^{-2}|a_1|^2 \right)^{-0.5} = \left( 1 + \left( |A_1|^{-2} - 1 \right) |a_1|^2 \right)^{-0.5}$. Results of the protocol are presented in Table 3.

| Measurement outcomes | Obtained states | Success probabilities |
|---|---|---|
| $(0,0)$, $(1,0)$ | $|\psi_{01}^{(out)}\rangle$ (AM) | $P_{01}(\alpha)$ |
| $(0,1)$, $(1,1)$ | $|\psi_{11}^{(out)}\rangle$ (original) | $P_{11}(\alpha)$ |

**Table 3**. Results of quantum teleportation protocol of initial AM unknown qubit (46) in the case of $\alpha << 1$ when preparatory actions of Charles to prepare initial AM qubit from original unknown one are not involved in the protocol. Bob can directly restore the original qubit with probability $P_{11}(\alpha)$ and the following relation $P_{01}(\alpha) + P_{11}(\alpha) \approx 1$ is performed in the case.

Here, we use the following notations

$$|\psi_{11}^{(out)}\rangle = \left| \begin{array}{c} a_0 \\ a_1 \end{array} \right|, \qquad (47)$$

$$|\psi_{n1}^{(out)}\rangle = N_{n1} \left| \begin{array}{c} a_0 \\ A_n A_1^{-1}a_1 \end{array} \right|, \qquad (48)$$

where the normalization factor is $N_{n1} = \left( |a_0|^2 + |A_n|^2|A_1|^{-2}|a_1|^2 \right)^{-0.5} = \left( 1 + \left( |A_n|^2|A_1|^{-2} - 1 \right) |a_1|^2 \right)^{-0.5}$ and $n \neq 1$. The success probabilities are the following

$$P_{11}(\alpha) = \exp\left(-|\alpha|^2\right) |\alpha|^2 N_1^{(in)2} = \frac{\exp\left(-|\alpha|^2\right)|\alpha|^2}{1 + \left( |A_1|^{-2} - 1 \right)a_1|^2}, \qquad (49)$$

$$P_{n1}(\alpha) = \exp\left(-|\alpha|^2\right) \frac{|\alpha|^{2n}}{n!} \frac{N_1^{(in)2}}{N_{n1}^2} = \exp\left(-|\alpha|^2\right) \frac{|\alpha|^{2n}}{n!} \frac{1 + \left( |A_n|^2|A_1|^{-2} - 1 \right)a_1|^2}{1 + \left( |A_1|^{-2} - 1 \right)a_1|^2}, \qquad (50)$$

where $n = 0$ is taken in Table 3. This third distribution is normalized $\sum_{n=0}^{\infty} P_{n1}(\alpha) = 1$ and different from two others (37, 44,45). Plots of the probabilities $P_{01}$, $P_{11}$ and $P_{21}$ in dependency on $|a_1|$ are shown in figures 5(a-d) for different values of the displacement parameter $\alpha$. As well as for the distribution (44,45), there is the range of amplitude values $|a_1|$ for which the teleportation of unknown qubit (10) occurs with a success probability greater than $0.5$. This range is shifted to amplitude values $|a_1| \cong 1$ in contrast to the case of amplitude modulation (41).

## 5 Amplitude demodulation and increase in efficiency

In the previous section, we considered the possibility for Alice to teleport AM unknown qubit to Bob provided that Charles has supplied her by the states and, at the same time, the Charles's actions are not included in the calculation of quantum teleportation. Technique of initial amplitude modulation of an unknown qubit allows us to implement quantum teleportation protocol with probability of success more than $0.5$ but only for two cases, when the teleported qubit is significantly unbalanced either with $|a_1| < 0.4$ (Figs. 4) or $|a_1| > 0.95$ (Figs. 5). Here, the problem of amplitude demodulation becomes relevant for one of the states (either $\left| \psi_{10}^{(out)} \right\rangle$ or $\left| \psi_{01}^{(out)} \right\rangle$) in order to increase success probability of the protocol unlike the case of direct teleportation (Table 1). Nevertheless, this strategy with the initial amplitude modulation is more preferable compared with the case discussed in the previous section as it guarantees exact teleportation with some probability and decreases number of the states requiring amplitude demodulation (one state instead of two). Moreover, if the teleported qubit is highly unbalanced, then the success probability of the protocol may become close to one provided that Charles successfully guessed with amplitude modulation of the qubit. Note also that Bob unambiguously knows which of two states he obtained either original (42,47) or AM (43,48) after Alice has sent him auxiliary classical information in length of 2 bits.

Consider for Bob two possible options to implement demodulation of obtained qubits. One of them is related with interaction of strong coherent field with AM qubit on HTBS as shown in Appendix B and another is based on quantum swapping [37]. First consider the case in which Charles realizes the state (41) and hands it to Alice who teleports it according to the protocol examined. Suppose that Bob, having received an unknown qubit, demodulates it by mixing it with a strong coherent field on HTBS as shown in Fig. 7(a) with coherent state as input. For this reason, we call this method coherent. The mathematical details of this method are given in Appendix B (B1-B4). Then, he obtains original unknown qubit with success probability

$$P_0^{(C)}(\alpha) = \frac{\exp\left(-|\alpha|^2\right)}{1 + \left(|A_0|^{-2} - 1\right)|a_1|^2} \left(1 + \exp\left(-|\gamma_1|^2\right)|\alpha|^2 \left|1 - |\gamma_1|^2\right|^2\right), \tag{51}$$

where the parameter $\gamma_1$ is determined from the condition $A_1(\alpha) A_0^{-1}(\alpha) c_{01}(\gamma_1) = c_{11}(\gamma_1)$ and can be estimated as $\gamma_1 \approx \alpha^2$ in the case of $\alpha \ll 1$. Here, the superscript $C$ implies that Bob made use of coherent method of interaction of AM unknown qubit with strong coherent state on HTBS. If Charles transmits AM unknown qubit (46) to Alice, then Bob can extract original state with success probability

$$P_1^{(C)}(\alpha) = \frac{\exp\left(-|\alpha|^2\right)|\alpha|^2}{1 + \left(|A_1|^{-2} - 1\right)|a_1|^2} \left(1 + \exp\left(-|\gamma_2|^2\right)|\gamma_2|^2 / |\alpha|^2\right), \tag{52}$$

where the parameter $\gamma_2$ is determined from the condition $A_0(\alpha) A_1^{-1}(\alpha) = c_{10}(\gamma_2)$ and can be also estimated as $\gamma_2 \approx \alpha^2$ in the case of $\alpha \ll 1$. Note only that there are still two AM unknown states left at Bob's disposal in the base $\{|0\rangle, |1\rangle\}$ that he identifies

$$\left|\psi_{10}^{(out)'}\right\rangle_3 = N_{10}'\left(a_0|0\rangle_3 + A_1(\alpha)A_0^{-1}(\alpha)A_0^{-1}(\gamma_1)a_1|1\rangle_3\right), \tag{53a}$$

$$\left|\psi_{01}^{(out)'}\right\rangle_3 = N_{01}'\left(a_0|0\rangle_3 + A_0(\alpha)A_1^{-1}(\alpha)A_1^{-1}(\gamma_2)a_1|1\rangle_3\right), \tag{53b}$$

where $N_{10}'$ and $N_{01}'$ are the corresponding normalization coefficients. He can locally continue the procedure for demodulating AM unknown states (53a) and (53b) with aim to increase the success probabilities (51,52) to restore original qubit.

Consider another possibility for Bob to extract original unknown qubit from obtained AM ones by quantum swapping with known qubits [37]. So, Bob mixes the state $\left|\psi_{10}^{(out)}\right\rangle_{12}$ (Eq. (43)) with the following auxiliary state

$$N_0'\left(A_1A_0^{-1}|01\rangle_{34} + |10\rangle_{34}\right), \tag{54a}$$

and the state $\left|\psi_{01}^{(out)}\right\rangle_{12}$ (Eq. (48)) with the state

$$N_1'\left(A_0A_1^{-1}|01\rangle_{34} + |10\rangle_{34}\right), \tag{54b}$$

on balanced beam splitter. Here, the normalization factors $N_0' = \left(|A_1|^2|A_0|^{-2} + 1\right)^{-0.5}$ and $N_1' = \left(|A_0|^2|A_1|^{-2} + 1\right)^{-0.5}$ are introduced. Note only Bob mixes second and third modes of the states. Then, Bob obtains the original unknown qubit in base $\left\{|01\rangle, |10\rangle\right\}$ with total success probability

$$P_0^{(S)}(\alpha) = \frac{\exp\left(-|\alpha|^2\right)}{1 + \left(|A_0|^{-2} - 1\right)|a_1|^2}\left(1 + \frac{|\alpha|^2\left|1 - |\alpha|^2\right|^2}{|\alpha|^4 + \left|1 - |\alpha|^2\right|^2}\right), \tag{55}$$

provided that he registered the following events $|01\rangle_{23}$ and $|10\rangle_{23}$ in first case (Eq. (54a)) and

$$P_1^{(S)}(\alpha) = \frac{\exp\left(-|\alpha|^2\right)|\alpha|^2}{1 + \left(|A_1|^{-2} - 1\right)|a_1|^2}\left(1 + \frac{|\alpha|^2}{|\alpha|^4 + \left(1 - \alpha^2\right)^2}\right). \tag{56}$$

in second case (Eq. (54b)). Superscript $S$ is responsible for swapping operation. Here the protocol of the quantum teleportation ends as Bob loses all control over the other AM qubits.

Corresponding plots of the success probabilities $P_0^{(C)}(\alpha)$ (Eq. (51)), $P_1^{(C)}(\alpha)$ (Eq. (52)) and $P_0^{(S)}(\alpha)$ (Eq. (55)), $P_1^{(S)}(\alpha)$ (Eq. (56)) are shown in figures 6(a,b), respectively. As can be seen from these graphs, these dependencies are of more interest from the point of view of increasing the efficiency of the protocol compared with the case when Bob made no efforts to demodulate the AM qubit as shown in Figs. 4,5. As well as in Figures 4 and 5, there are regions in these dependencies in which the success probabilities can reach values at least greater than $0.5$. But such areas are significantly larger than those in Figures 4 and 5. This confirms the view that own Bob's efforts to demodulate an unknown qubit can lead to an increase of the success probability of the protocol. As well as for the cases in Figures 4 and 5, the probability of success for Bob to get the original qubit becomes close to unity in the case of an unbalanced qubit. But an increase of the contribution of Bob's efforts to demodulate the AM qubits to the total success probability is particularly pronounced with an increase of the displacement amplitude that may require use of a superconducting single-photon detector [38] (SSPD) working at cryogenic temperature to recognize the state with more number of photon in the teleported mode. It is quite possible that SSPD are also needed to recognize even and odd SCS when their amplitudes growing. If Charles has access to the amplitude information about the teleported state not knowing anything about the phase relations of the qubit, he can choose the relevant modulation factor in order to ensure the greatest possible success probability of the quantum teleportation of the unknown qubit. Note that even partial knowledge about qubit (information about amplitudes) leaves the qubit unknown. Numerical calculations show that the probabilities of success $P_0^{(C)}(\alpha)$, $P_0^{(S)}(\alpha)$ $\left(P_0^{(C)}(\alpha) \approx P_0^{(S)}(\alpha)\right)$ and $P_1^{(C)}(\alpha)$, $P_1^{(S)}(\alpha)$ $\left(P_1^{(C)}(\alpha) \approx P_1^{(S)}(\alpha)\right)$ behave approximately identically.

This means that there is no particular preference what kind of demodulation strategy to choose by Bob, provided that he no longer makes any efforts. But a strategy with mixing with a coherent pump wave can improve the success probability of success for Bob provided he continues his efforts on demodulation of obtained states (53a) and (53b). Consider one of the possibilities. Let Bob pass the state $\left|\psi_{10}^{(out)'}\right\rangle_3$ (53a) through the beam splitter with $t \to 0$ and $r \approx 0$ with further registration of vacuum in auxiliary mode. Then, Bob again obtains the original state (10) with success probability

$$\delta P_0^{(C)}(\alpha) = \frac{\exp\left(-|\alpha|^2\right)|\alpha|^2}{1 + \left(|A_0|^{-2} - 1\right)|a_1|^2} \exp\left(-|\gamma_1|^2\right)|\gamma_1|^2 , \tag{57}$$

which is additionally added to expression (51). Contribution of the term becomes significant with increase of the displacement amplitude $\alpha$.

Let us involve Charles actions with unknown qubit in calculation of success probabilities of the teleportation protocol in the light of comparing it with value 0.5 that is observed in the Bell state formalism [8]. Assume that Charles can generate states (41,46) from original unknown qubit (10) with probabilities $p_0$ and $p_1$, respectively, with condition $p_0 + p_1 = 1$ or, at least, $p_0 + p_1 \approx 1$. The probabilities can also include defects in communication lines in the transfer of the states from Charles to Alice. Then, the total success probability for Bob to restore original qubit involving preparatory actions of Charles over the qubit is given by

$$P_{tot}^{(C)}(\alpha) = p_0 P_0^{(C)}(\alpha) + p_1 P_1^{(C)}(\alpha), \tag{58a}$$

$$P_{tot}^{(S)}(\alpha) = p_0 P_0^{(S)}(\alpha) + p_1 P_1^{(S)}(\alpha). \tag{58b}$$

In particular, we have either $P_{tot}^{(C)}(\alpha) = P_0^{(C)}(\alpha)$, $P_{tot}^{(S)}(\alpha) = P_0^{(S)}(\alpha)$ in the case of $p_0 = 1$, $p_1 = 0$ or $P_{tot}^{(C)}(\alpha) = P_1^{(C)}(\alpha)$, $P_{tot}^{(S)}(\alpha) = P_1^{(S)}(\alpha)$ in the case of $p_0 = 0$, $p_1 = 1$. Here, it is possible to discuss various strategies for preparing AM unknown qubits by Charles. Charles can choose in advance a certain strategy with constant values $p_0$ and $p_1$ independent of the parameters of the teleported qubit, for example $p_0 = p_1 = 0.5$. Such a strategy can assume that some qubits can be discarded in process of their transformation. Suppose that Charles knows the information that the absolute value of the teleported qubit $|a_1| \approx 0$ without knowing anything about its phase information. Then it is natural to assume that Charles should produce the state (41) by discarding all the others to ensure $p_0 = 1$, $p_1 = 0$. The opposite case takes place in the case $|a_1| \approx 1$. It is quite possible that the quantities $p_0$ and $p_1$ may also become dependent functions of $|a_1|$. Then it is necessary to ensure the condition that $p_0$ and $p_1$ tend to the maximum possible values when $|a_1| \approx 0$ and $|a_1| \approx 1$, respectively. Let us discuss one of possible possibilities for Charles practically to gain access to the amplitude information about a qubit. He can begin by passing a single photon through a beam splitter with corresponding known parameters. After that, he can use the procedure of interaction of this qubit with coherent state on HTBS described in Appendix B. Finally, Charles can get AM qubit with known amplitude parameters and an unknown phase difference imposed by the coherent state, provided that a certain event is registered (Fig. 7(a)).

It is interesting to consider Bob's state after Alice has performed measurement but before Bob has learned the measurement results. Consider it on example of direct implementation of the protocol (Table 1). Then, using the formulas (34a), (34b), (37), (A22), it is possible to show that Bob obtains the state

$$\rho_B = \frac{1}{2}\left(|01\rangle\langle01| + |10\rangle\langle10|\right) + \frac{\exp\left(-2|\alpha|^2\right)\exp\left(-2|\beta|^2\right)}{2}$$

$$\left(\begin{array}{l}\left(|a_0|^2 + \left(1-4|a_1|^2\right)|a_1|^2 - 2\alpha^* a_0^* a_1 + 2\alpha a_0 a_1^*\right)|01\rangle\langle10| + \\ \left(|a_0|^2 + \left(1-4|a_1|^2\right)|a_1|^2 - 2\alpha a_0 a_1^* + 2\alpha^* a_0^* a_1\right)|10\rangle\langle01|\end{array}\right). \tag{59}$$

In the case studied, the resulting state has off-diagonal terms dependent on the parameters of the teleported state. Although, it is worth noting that off-diagonal terms rapidly disappear with increasing the displacement amplitude when either $\alpha \to \infty$ or $\beta \to \infty$. We conjectured certain size $\beta$ of coherent components of the quantum channel (11). But it follows from Eqs. (8,9), it is only very good approximation. In exact consideration, size of coherent components must be infinite to implement displacement of arbitrary state that nullifies off-diagonal elements. Thus, the obtained state has no dependence on parameters of the teleported state preventing Alice from using the teleportation technique to transmit information to Bob faster than speed of light as in four-dimensional space with Bell states [7]. Note the quantum channel is not maximally entangled. It becomes maximally entangled in the case displacement amplitude approaching to infinity.

Suppose that Alice should teleport to Bob two orthogonal qubits

$$|\varphi_1\rangle_2 = a_0|0\rangle_2 + a_1|1\rangle_2, \tag{60a}$$

$$|\varphi_2\rangle_2 = a_1^*|0\rangle_2 - a_0^*|1\rangle_2, \tag{60b}$$

where the orthogonality condition $\langle\varphi_1|\varphi_2\rangle = 0$ is performed. Assume that Alice choose amplitude modulation with amplitude factor $A_0^{-1}$ as in (41) for the teleported states (60a) and (60b). Performing the procedure as described above, Bob, in particular, receives the same original qubits (60a) and (60b) with corresponding success probability (44). Let's assume that $|a_1| \ll 1$ and Alice knows about this circumstance and about what qubits she has either (60a) and 60(b). Then, she needs to modulate the qubit (60a) by the factor $A_0^{-1}$ and state (60b) by the factor $A_1^{-1}$ for Bob to obtain output orthogonal qubits with success probabilities close to one.

## 6 Results

We considered mechanism of interaction of continuous variable states with discrete variable states (continuous-discrete interaction) on HTBS to implement quantum teleportation protocol of unknown qubit beyond of Bell states formalism [26]. Coherent components of the hybrid state (11) simultaneously displace unknown teleported qubit in indistinguishable manner on HTBS by the values that differ from each other only by sign $\pm\alpha$. Both coherent components of the quantum channel displace the teleported state so that we do not have access to information on what value (either $\alpha > 0$ or $\alpha < 0$) the teleported qubit is displaced despite the fact that one of these displacements has already happened. Projection of the uncertain teleported state being superposition of components from different Hilbert spaces determined by parameters $\alpha$ and $-\alpha$ (A3) onto measurement basis (A4) produces desired controlled $-Z$ operation due to relation (A13) and teleported state can be partly recovered. The same mechanism is applied for generation of the quantum channel (11). The teleported states become amplitude modulated in direct realization of the protocol. Amplitude modulation or amplitude distortion of the output state on compared with original, at least from a mathematical point of view, arises due to displaced vacuum (coherent state) and displaced single photon are transformed differently from each other under projection onto measurement basis. Amplitude modulation of the output states is an inalienable and inherent feature of the protocol as an inability to implement complete Bell-state measurement by linear optics methods [8]. In proposed implementation, Alice can distinguish all her measurement outcomes and Bob knows exactly what state he has in his hands. Limit in 50 percent of success probability to distinguish the states imposed by linear optics for Bell states is not relevant to the proposed scheme.

The effectiveness of this approach depends on the strategy that the participants of the protocol can choose. We considered one of the possible strategies with the third member who helps to prepare input AM unknown qubits for Alice teleportation. In this strategy, preparing unknown AM qubit from the initial is a preliminary operation and it is not included in the protocol of quantum teleportation. The use of the consideration is appropriate [9,10]. Then, the success probability becomes a function of the absolute value of the amplitude of the teleported qubit. If Charles exactly knows that the unknown qubit is strongly unbalanced, he can make efforts to create appropriate AM qubit and hand it to Alice for teleportation with large success probability. Moreover, such teleportation can be made in a realistic scenario with realizable quantum channel of small amplitude and measurement of the number of photons and the parity of the SCS states can be replaced by on-off measurement (there is or no photon in mode). This realization with two commercially achievable APDs significantly reduces the technical difficulties in the practical implementation of the protocol. SCS of such size can be produced in practice [27-29,39].

A more complicated case occurs if the unknown qubit is clearly balanced or if Charles does not have any information about the teleported qubit. Here, Bob's efforts to demodulate the obtained AM qubits can play an important role to increase success probability. We have considered only two possible methods for Bob to demodulate his qubits. It should be especially noted that both amplitude modulation and demodulation are carried out in a controlled manner and the protocol participants (Alice and Bob) do not lose any control over the qubits during all process, in contrast to Bell states measurement when outcomes of two states are not distinguishable [8-10]. Bob's efforts to demodulate unknown qubits lead to an increase in the success probability, especially with an increase of the displacement amplitude which may impose more serious conditions on measurements. It is also interesting to consider other possibilities (most likely nonlinear) in order to demodulate unknown qubits. Despite the fact that our proposal produces high fidelity quantum teleportation for a restricted set of qubits, it does not require auxiliary photons, complex quantum channels and hyperentanglement and can be implemented with irreducible number of optical elements: one beam splitter and two APDs unlike [40].

## Appendix A. Properties of the displacement operator under change of its amplitude on opposite in sign

Unitary displacement operator [34] is determined by

$$D(\alpha) = \exp\left(\alpha a^+ - \alpha^* a\right), \tag{A1}$$

where $\alpha$ is an amplitude of the displacement and $a$, $a^+$ are the bosonic annihilation and creation operators. Its action on number state results in

$$|n, \alpha\rangle = D(\alpha)|n\rangle, \tag{A2}$$

where the same notations as in [23] are used. The displaced number states (A2) are defined by two numbers: quantum discrete number $n$ and classical continuous parameter $\alpha$ which can be recognized as their size [23]. The states (A2) belong to vector (Hilbert) space with appropriate inner product $\langle n, \alpha | m, \alpha \rangle = \delta_{nm}$ with $\delta_{nm}$ being Kronecker delta [26]. The Hilbert space is determined by the displacement amplitude $\alpha$ with the base states

$$\left\{|n, \alpha\rangle, n = 0, 1, 2, ..., \infty\right\}. \tag{A3}$$

If the displacement amplitude is $\alpha = 0$, then we deal with Hilbert space of the number states

$$\left\{|n\rangle, n = 0, 1, 2, ..., \infty\right\}. \tag{A4}$$

As the set of the states (A3) is complete [23], any displaced number state (A2) from different Hilbert space can be represented in the terms of the base states. So, the number state and their displaced counterparts are related with each other as

$$|l,\alpha\rangle = F\sum_{n=0}^{\infty} c_{\ln}(\alpha)|n\rangle, \tag{A5}$$

where multiplier $F = \exp\left(-|\alpha|^2/2\right)$ is introduced and the matrix elements are the following [23]

$$c_{\ln}(\alpha) = \frac{\alpha^{n-l}}{\sqrt{l!}\sqrt{n!}}\sum_{k=0}^{l}(-1)^k C_l^k |\alpha|^{2k}\prod_{k=0}^{l-1}(n-l+k+1), \tag{A6}$$

or the same

$$c_{\ln}(\alpha) = \frac{\alpha^{n-l}}{\sqrt{l!}\sqrt{n!}}\begin{pmatrix} n(n-1)...(n-l+1)-C_l^1|\alpha|^2 n(n-1)...(n-l+2)+ \\ C_l^2|\alpha|^4 n(n-1)...(n-l+3)+...+(-1)^k C_l^k|\alpha|^{2k}\prod_{k=0}^{l-1}(n-l+k-1)+ \\ (-1)^l|\alpha|^{2l} \end{pmatrix}, \tag{A7}$$

where $C_l^k = l!/\left(k!(l-k)!\right)$ are the elements of the Bernoulli distribution and number product is

$$\prod_{k=0}^{l-1}(n-l+k+1) = n(n-1)...(n-l+1). \tag{A8}$$

It is worth noting that the reverse transformation $c_{\ln}(-\alpha)$ defines number state through their displaced analogies due to unitary nature of the displacement operator (A1). Consider partial cases with $l=0$ and $l=1$ corresponding to expression of coherent state and displaced single photon in the terms of the number states [23]

$$|0,\alpha\rangle = F\sum_{n=0}^{\infty} c_{0n}(\alpha)|n\rangle, \tag{A9}$$

$$|1,\alpha\rangle = F\sum_{n=0}^{\infty} c_{1n}(\alpha)|n\rangle, \tag{A10}$$

with the matrix elements

$$c_{0n}(\alpha) = \frac{\alpha^n}{\sqrt{n!}}, \tag{A11}$$

$$c_{1n}(\alpha) = \frac{\alpha^{n-1}}{\sqrt{n!}}\left(n-|\alpha|^2\right). \tag{A12}$$

The matrix elements (A11,A12) directly stems from general formulas (A6,A7).

The matrix elements (A6,A7) consist of a common factor proportional to $\alpha^{n-l}$ and a polynomial expression of degree $l$ being the maximum of the degree of its monomial over $|\alpha|^2$ which is enclosed in parentheses. Thus, the term $\alpha^{n-l}$ defines the behavior of the matrix elements under change of the displacement amplitude on opposite in sign that leads to

$$c_{\ln}(-\alpha) = (-1)^{n-l} c_{\ln}(\alpha). \tag{A13}$$

In particular, we have the following rules for decomposition of the even displaced number states $l=2m$

$$c_{2mn}(-\alpha) = (-1)^n c_{2mn}(\alpha), \tag{A14}$$

and of the odd displaced number states $l=2m+1$

$$c_{2m+1n}(-\alpha) = (-1)^{n-1} c_{2m+1n}(\alpha). \tag{A15}$$

In application to the matrix elements of coherent and displaced single photon states, we have

$$c_{0n}(-\alpha) = (-1)^n c_{0n}(\alpha), \tag{A16}$$

$$c_{1n}(-\alpha) = (-1)^{n-1} c_{1n}(\alpha). \tag{A17}$$

The probability distributions of vacuum and single photon over number states displaced on arbitrary value $\alpha$ are defined by

$$P_{0n}(\alpha) = F^2 |c_{0n}(\alpha)|^2 = \exp(-|\alpha|^2) \frac{|\alpha|^{2n}}{n!}, \tag{A18}$$

$$P_{1n}(\alpha) = F^2 |c_{10n}(\alpha)|^2 = \exp(-|\alpha|^2) \frac{|\alpha|^{2(n-1)}}{n!} (n - |\alpha|^2)^2, \tag{A19}$$

respectively. It is possible directly to check the matrix elements $c_{mn}(\alpha)$ (A6,A7) satisfy the normalization condition [23]

$$F^2 \sum_{n=0}^{\infty} |c_{mn}(\alpha)|^2 = F^2 \sum_{n=0}^{\infty} |c_{mn}(-\alpha)|^2 = 1. \tag{A20}$$

The normalization condition for vacuum $\sum_{n=0}^{\infty} P_{0n}(\alpha) = \sum_{n=0}^{\infty} P_{0n}(-\alpha) = 1$ and single photon $\sum_{n=0}^{\infty} P_{1n}(\alpha) = \sum_{n=0}^{\infty} P_{1n}(-\alpha) = 1$ can be directly checked using (A11,A12). Due to orthogonality of the displaced states with different numbers $n$, we have

$$F^2 \sum_{n=0}^{\infty} c_{mn}^*(\alpha) c_{kn}(\alpha) = \delta_{mk}. \tag{A21}$$

In particular, we have

$$F^2 \sum_{n=0}^{\infty} c_{0n}^*(\alpha) c_{1n}(\alpha) = 0. \tag{A22}$$

### Appendix B. Amplitude demodulation and generation of the hybrid quantum channel

Consider one of the possibilities for Bob to demodulate AM states either $\left|\psi_{10}^{(out)}\right\rangle$ (43) or $\left|\psi_{01}^{(out)}\right\rangle$ (48) by its interaction with strong coherent state on HTBS in Fig. 7(a). Consider it on example of the state (43). Then, we have the following

$$BS_{13}\left(\left|0,-\gamma_1\right\rangle_1 \left|\psi_{10}^{(out)}\right\rangle_{23}\right) \to FN_{10}\left|0,-\gamma_1\right\rangle_1 c_{10}(\gamma_1 \left(a_0\left|0\right\rangle_2 + A_1 A_0^{-1}\frac{c_{00}(\gamma_1)}{c_{10}(\gamma_1)} a_1\left|1\right\rangle_3\right)\left|0\right\rangle_3 +$$
$$FN_{10}\left|0,-\gamma_1\right\rangle_1 c_{11}(\gamma_1 \left(a_0\left|0\right\rangle_2 + A_1 A_0^{-1}\frac{c_{01}(\gamma_1)}{c_{11}(\gamma_1)} a_1\left|1\right\rangle_3\right)\left|1\right\rangle_3 \tag{B1}$$

in regime $\gamma_1 << 1$. If we adopt the condition

$$A_1 A_0^{-1}\frac{c_{01}(\gamma_1)}{c_{11}(\gamma_1)} = 1, \tag{B2}$$

then Bob obtains the original unknown qubit (10) provided that he registered the single photon in the second mode. If he fixed vacuum, then Bob obtains AM state (53a). Using the condition (B2), we can estimate the value of the amplitude $\gamma_1$ as given above. The considered method is applicable to demodulation of the state $\left|\psi_{01}^{(out)}\right\rangle$. Indeed, we have

$$BS_{13}\left(\left|0,-\gamma_2\right\rangle_1 \left|\psi_{01}^{(out)}\right\rangle_{23}\right) \to FN_{01}\left|0,-\gamma_2\right\rangle_1 c_{10}(\gamma_2 \left(a_0\left|0\right\rangle_2 + A_0 A_1^{-1}\frac{c_{00}(\gamma_2)}{c_{10}(\gamma_2)} a_1\left|1\right\rangle_3\right)\left|0\right\rangle_3 +$$
$$FN_{01}\left|0,-\gamma_2\right\rangle_1 c_{11}(\gamma_2 \left(a_0\left|0\right\rangle_2 + A_0 A_1^{-1}\frac{c_{01}(\gamma_2)}{c_{11}(\gamma_2)} a_1\left|1\right\rangle_3\right)\left|1\right\rangle_3 \tag{B3}$$

in regime $\gamma_2 << 1$. If we impose the following condition

$$A_0 A_1^{-1}\frac{c_{00}(\gamma_2)}{c_{10}(\gamma_2)} = 1, \tag{B4}$$

then Bob has at his disposal the original qubit (10) provided that he registered vacuum in the second mode. If the result of his measurement is a single photon, then he obtains the AM state (53b). We can calculate the value of $\gamma_2$ using (B4). The results are also used to calculate the total success probabilities of the teleportation (51) and (52) when Bob applies the technique of interaction with strong coherent state to demodulate his AM qubits.

The same mechanism of continuous-discrete interaction can be used to generate the hybrid state (11). Consider interaction of even SCS (33a) with maximally entangled state of two photons in superposition of two modes

$$|\phi\rangle_{3456} = |0101\rangle_{3456} + |1010\rangle_{3456},$$ (B5)

as shown in Fig. 7(b). The even SCS state interacts with fifth mode of the state (B5) on HTBS (1). Additionally, the coherent state $|-\beta_1\rangle_2$ occupying second mode interacts with its mode 6 on another HTBS

$$BS_{15}BS_{26}\left(|even\rangle_1|-\beta_1\rangle_2|\phi\rangle_{3456}\right) =$$
$$N_+\left(a_0 BS_{15}BS_{26}\left(|-\beta\rangle_1|-\beta_1\rangle_2|\varphi\rangle_{3456}\right) + a_1 BS_{15}BS_{26}\left(|\beta\rangle_1|-\beta_1\rangle_2|\varphi\rangle_{3456}\right)\right).$$ (B6)

Following the mathematical approach developed in section 2, we can obtain

$$BS_{15}BS_{26}\left(|even\rangle_1|\beta_1\rangle_2|\varphi\rangle_{3456}\right) \to \left|\Delta_1^{(id)}\right\rangle_{13456}|0,-\beta_1/t\rangle_2,$$ (B7)

in the limit case of $t=1$, $r=0$, where the ideal normalized state is the following

$$\left|\Delta_1^{(id)}\right\rangle_{13456} = N_+ F^2 \sum_{n=0}^{\infty}\sum_{m=0}^{\infty}|\Psi_{nm}\rangle_{134}|nm\rangle_{56},$$ (B8)

where the following states are introduced

$$|\Psi_{nm}\rangle_{134} = |0,-\beta\rangle_1\left|\varphi_{nm}^{(+)}(\alpha,\alpha_1)\right\rangle_{34} + (-1)^n|0,\beta\rangle_1\left|\varphi_{nm}^{(-)}(\alpha,\alpha_1)\right\rangle_{34},$$ (B9)

with

$$\left|\varphi_{nm}^{(\pm)}(\alpha,\alpha_1)\right\rangle_{34} = \left(c_{0n}(\alpha)c_{1m}(\alpha_1)|01\rangle_{34} \pm c_{1n}(\alpha)c_{0m}(\alpha_1)|10\rangle_{34}\right)/\sqrt{2}.$$ (B10)

Deterministic displacement of the state (B5) by the values $\pm\alpha$ by the coherent components of even SCS with disappearing all information about the events is followed by a probabilistic measurement in auxiliary modes 5 and 6 in Fig. 6. Let us consider the case $\alpha = \alpha_1$. We are interested in registration of events either $|01\rangle_{56}$ or $|01\rangle_{56}$ that gives birth to the states

$$|\Psi_{01}\rangle_{134} = \left(|0,-\beta\rangle_1\left(|01\rangle_{34}+|10\rangle_{34}\right) + |0,-\beta\rangle_1\left(|01\rangle_{34}-|10\rangle_{34}\right)\right)/\sqrt{2},$$ (B11)

and

$$|\Psi_{10}\rangle_{134} = \left(|0,-\beta\rangle_1\left(|01\rangle_{34}+|10\rangle_{34}\right) - |0,-\beta\rangle_1\left(|01\rangle_{34}-|10\rangle_{34}\right)\right)/\sqrt{2}.$$ (B12)

Application of Hadamard gate (36a) with subsequent $Z^n$ (36b), where $n=0,1$, generates the input channel (11). It can be shown that, as the case of the quantum teleportation protocol, the probability of the measurement outcomes $(00)$, $(01)$, $(10)$ and $(11)$ significantly prevails over higher order measurement events $(mn)$ in the case of $\alpha << 1$ which makes possible to use two APD (Fig. 7b). The method of generation of the hybrid entangled state (11) resembles approach used in [24] to experimentally generate another hybrid entangled state.

### Acknowledgement
The work was supported by Act 211 Government of the Russian Federation, contract № 02.A03.21.0011.

**List of figures**

**Figure 1**

A schematic representation of the quantum teleportation protocol of unknown qubit. Alice and Bob share hybrid entangled state (11). Alice has unknown qubit (10) which she wants to teleport to Bob. Alice mixes her part of the hybrid entangled with unknown qubit on HTBS with subsequent measurement in first coherent mode and second teleported mode by commercially achievable APDs in the case of $\alpha \ll 1$. After that she sends the result of her measurement to Bob and he performs unitary transformations ($Z$ and $H$ operations) on his part of the quantum channel. This scheme with the participation of Charles is assigned to increase the efficiency of the protocol in terms for Bob to obtain original (not AM) qubit. Charles performs preliminary actions on the initial amplitude modulation of the original unknown qubit. The Bob's demodulation efforts are also heeded in order to obtain the final success probabilities either $P_{0,1}^{(C)}$ (51,52) or $P_{0,1}^{(S)}$ (55,56) in dependency on demodulation strategy. Charles can prepare the initial AM unknown qubits $\left| \varphi_0^{(in)} \right\rangle$ (41) and $\left| \varphi_0^{(in)} \right\rangle$ (46) with probabilities $p_0$ and $p_1$, respectively. SHC means source of hybrid channel. AM means Charles's actions to modulate original unknown qubit. Demodulation procedures mean that which are offered in the work.

**Figure 2**

Dependency of fidelity *Fid* (Eq. (25)) on the displacement amplitude $\alpha$ and transmittance $t$ for input qubit (10) with amplitudes $a_0 = \sqrt{0.5}$, $a_1 = i\sqrt{0.5}$.

**Figure 3(a-e)**

Three-dimensional plots of probabilities (37) (a) $P_0$, (b) $P_1$, (c) $P_2$ and (d) $P_3$, respectively, in dependency on the displacement amplitude $\alpha$ and absolute value of amplitude $|a_1|$ of unknown qubit (10). Two-dimensional dependencies (e) show the probabilities $P_0$ (curve 1), $P_1$ (curve 2) and their sum $P_0 + P_1 \cong 1$ (curve 3) for $\alpha = 0.03$.

**Figure 4(a-d)**

Plots of dependencies of probabilities $P_{00}$, (curve 1) $P_{10}$, (curve 2) and $P_{20}$ (curve 3) (Eqs. (44, 45)) on $|a_1|$. The curves correspond to amplitude modulation of unknown qubit (41) previously performed by Charles. Charles actions are not involved in calculation of the probabilities. The plots are made for the following values of the displacement amplitude (a) $\alpha = 0.06$, (b) $\alpha = 0.1$, (c) $\alpha = 0.2$ and (d) $\alpha = 0.3$.

**Figure 5(a-d)**

Plots of dependencies of probabilities $P_{01}$ (curve 2), $P_{11}$ (curve 1) and $P_{21}$ (curve 3) (Eqs. (49,50)) on $|a_1|$. The curves correspond to another used amplitude modulation of unknown qubit (46) which is realized by Charles. Charles actions are not taken into account for calculation of the probabilities. The plots are made for the following values of the displacement amplitude (a) $\alpha = 0.06$, (b) $\alpha = 0.1$, (c) $\alpha = 0.2$ and (d) $\alpha = 0.3$.

**Figure 6(a,b)**

Plots of dependencies of the success probabilities (a) $P_0^{(C)}(\alpha)$ (Eq. (51)), $P_1^{(C)}(\alpha)$ (Eq. (52)) and (b) $P_0^{(S)}(\alpha)$ (Eq. (55)), $P_1^{(S)}(\alpha)$ (Eq. (56)) on $|a_1|$. Here, Charles efforts to construct AM state from original unknown qubit are not taken into calculations. The behavior of the curves depends on which modulation type is selected. The curves have increased values in regions $|a_1| << 1$ and $|a_1| \approx 1$ for the prepared AM states $\left|\varphi_0^{(in)}\right\rangle_2$ (41) and $\left|\varphi_1^{(in)}\right\rangle_2$ (46), respectively. The curves are made for (a) $\alpha = 0.1$ (curves 1 and 4), $\alpha = 0.3$ (curves 2 and 5) and $\alpha = 0.5$ (curves 3 and 6); (b) $\alpha = 0.2$ (curves 1 and 4), $\alpha = 0.4$ (curves 2 and 5) and $\alpha = 0.6$ (curves 3 and 6).

**Figure 7(a,b)**

(a) The optical scheme is used for amplitude demodulation of AM qubit performed by Bob. Bob mixes one mode of AM qubit with strong coherent state on HTBS with following registration of some measurement event that heralds about getting rid of AM qubit from amplitude factor. (b) Optical scheme used for generation of the hybrid entangled state (11). Additional interaction of auxiliary coherent state with entangled state (B5) is employed to complete the formation process of (11).

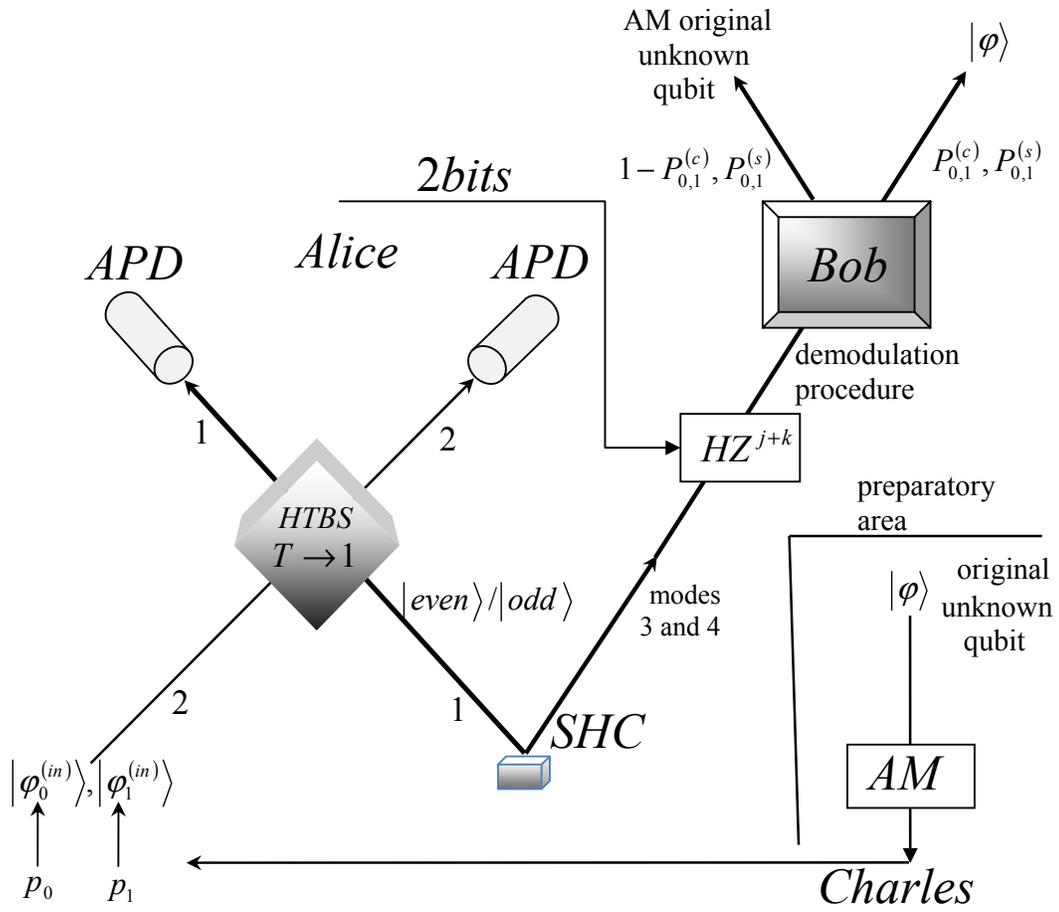

**Figure 1**

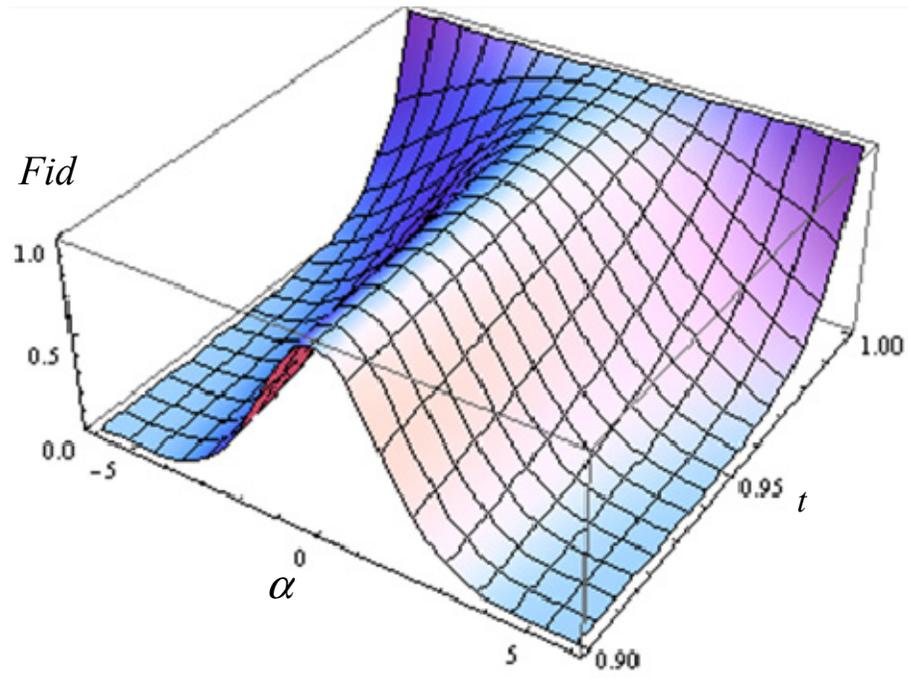

**Figure 2**

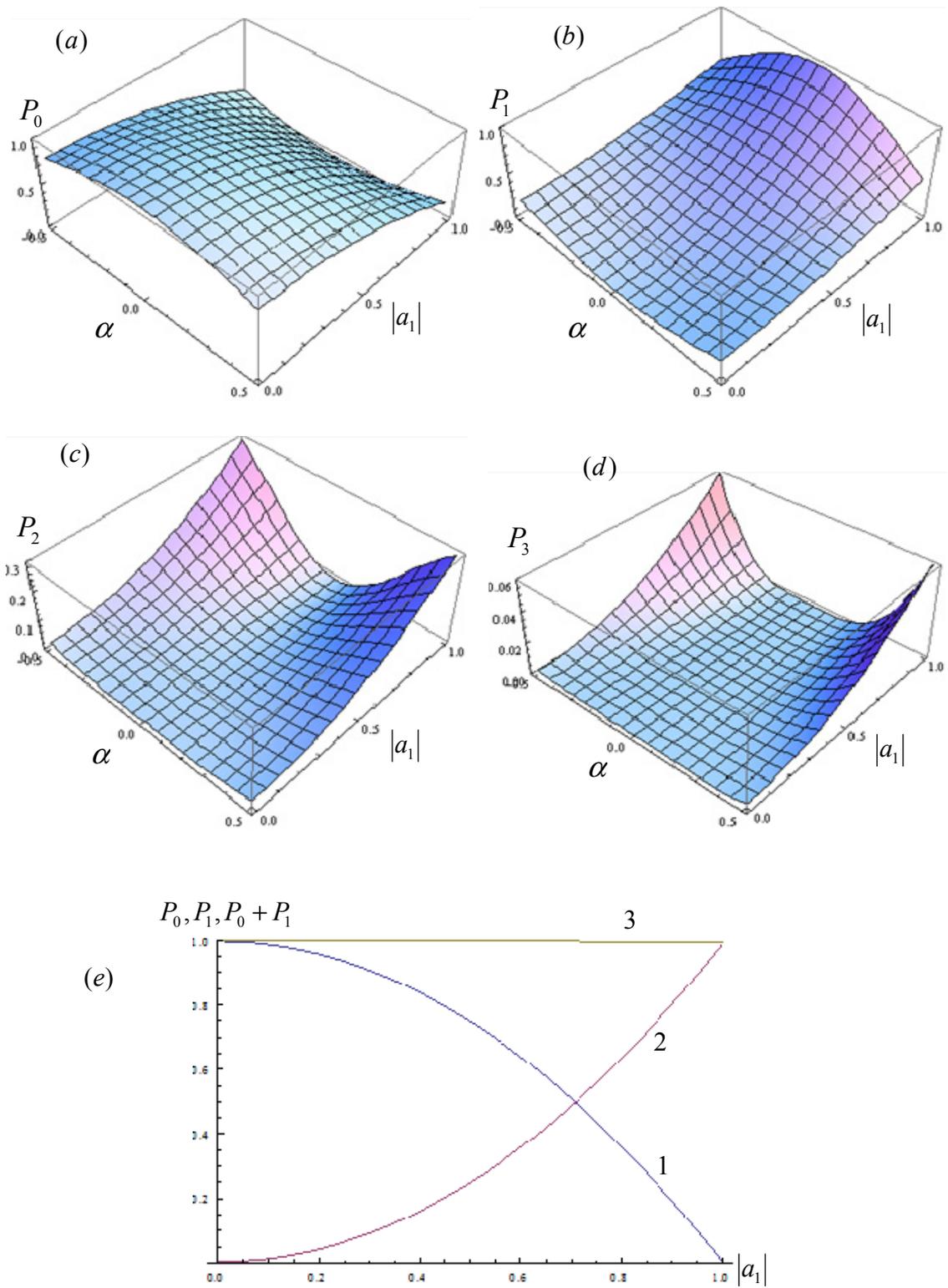

**Figure 3(a-e)**

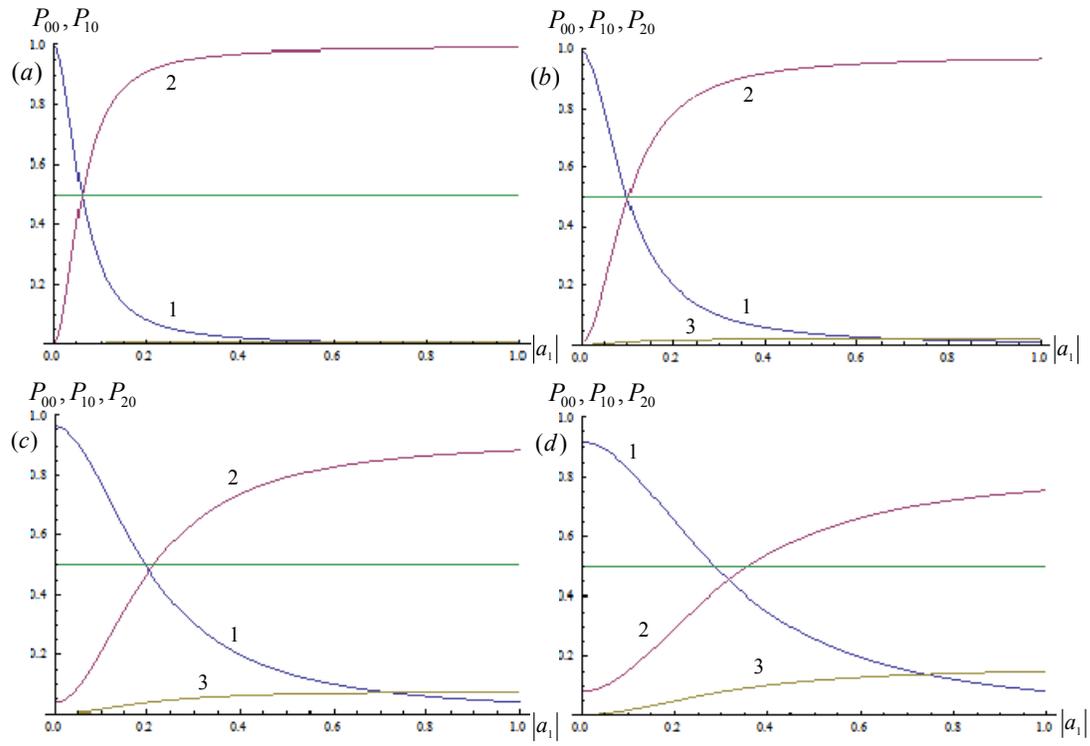

**Figure 4(a-d)**

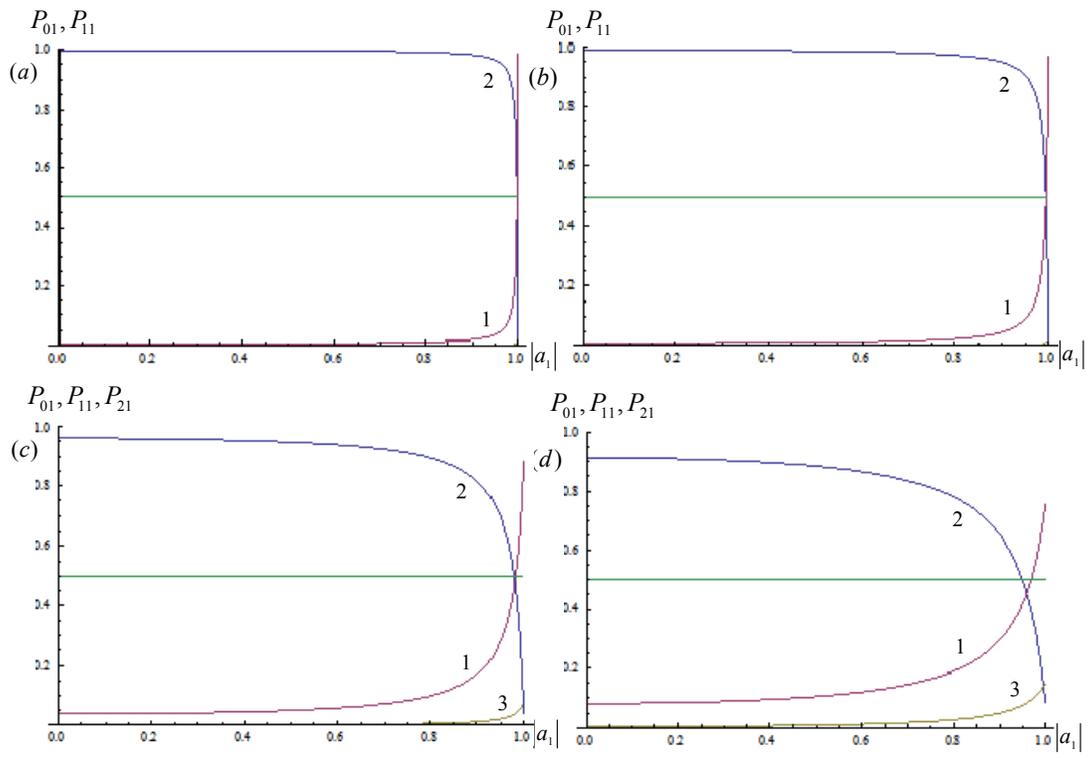

**Figure 5(a-d)**

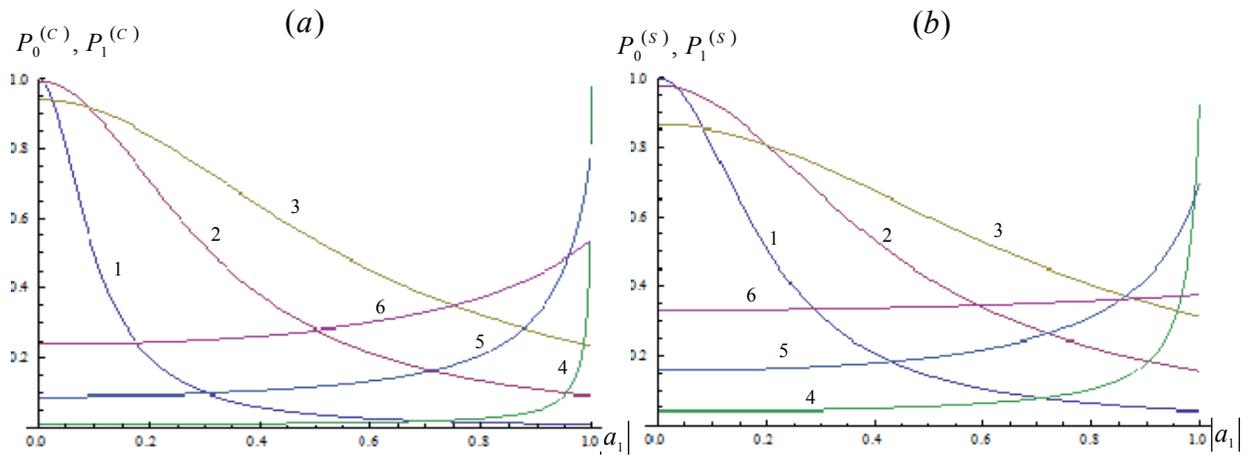

**Figure 6(a,b)**

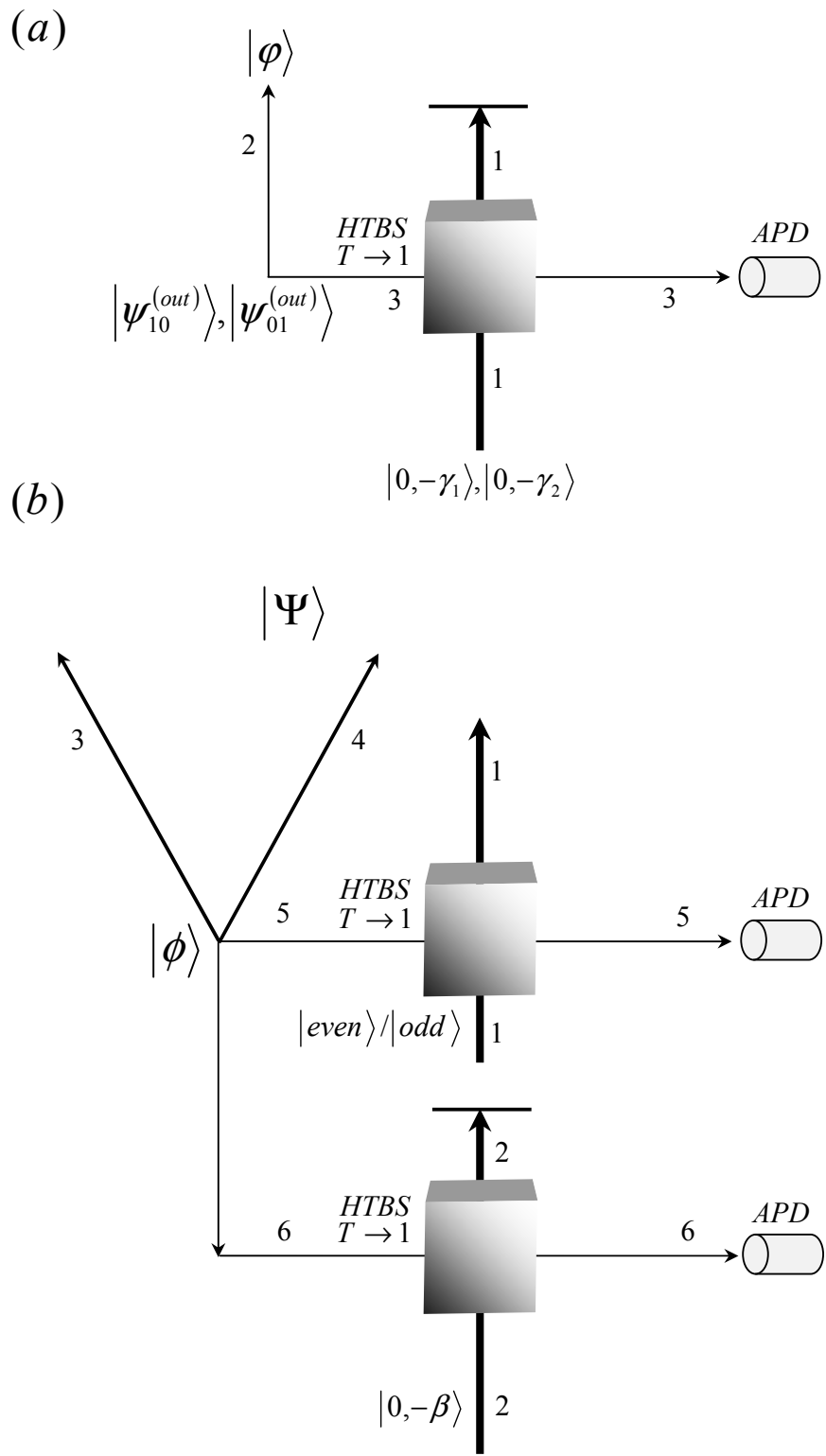

**Figure 7(a,b)**